\documentstyle[sprocl,epsf]{article}

\def\be{\begin{equation}}
\def\ee{\end{equation}}
\def\bea{\begin{eqnarray}}
\def\eea{\end{eqnarray}}

\begin{document}

\title{ Supersymmetric Left Right Model, Automatic R-parity Conservation 
and Constraints on the $W_R$ Mass}

\author{R. N. Mohapatra}

\address{Department of Physics, University of Maryland, College Park, 
Md-20742, USA\\E-mail: Rmohapat@physics.umd.edu}


\maketitle\abstracts{ Supersymmetric left-right models with the see-saw
mechanism for the neutrino masses have the attractive property that they
conserve baryon and lepton number exactly in the Lagrangian. In this talk,
I review the recent results valid 
for a large class of minimal versions of the model that supersymmetry
combined with the requirement that the ground state of the model conserve 
electric charge constrains the mass of the right 
handed $W_R$ boson to be in a certain range i.e. $M_{W_R}\leq$ 10 TeV or 
$\geq 10^{10}$ GeV. In the former case (low $M_{W_R}$), the vacuum 
breaks R-parity
spontaneously and the latter case (high $M_{W_R}$) is required if vacuum 
is to conserve R-parity.
In the second case, the effective low energy theory is the MSSM with
exact R-parity, nonvanishing neutrino masses and a pair of light doubly 
charged Higgs fields and their fermionic partners. Exact R-parity 
conservation via see saw mechanism therefore implies that the neutrino 
masses must be in the desired range to solve the solar and atmospheric 
neutrino puzzles.}

\section{Introduction}

Supersymmetry is now widely believed to be the next step beyond the 
successful standard model. Two primary reasons for this belief are:
(i) milder divergence structure of supersymmetry provides a way to 
maintain perturbative stability of the weak scale (or the Higgs mass)
and (ii) it also provides a mechanism to dynamically generate the
spontaneous breaking of the gauge symmetry through the use of the
renormalization group equations. Thus two of the major unsolved puzzles
of the standard model receive a rather satisfactory resolution. The minimal
supersymmetric model (MSSM) that leads to the standard model at low energies
provides the simplest realization of this idea and has been the subject 
of extensive investigation\cite{nath}. One of the key predictions of the MSSM
is the existence of a light neutral Higgs boson with mass less than 130 GeV
and can be used to test this model. And also another attractive feature
of the MSSM is that the lightest superpartner (LSP) of the standard model 
fields has all the right property to be the cold dark matter of the 
universe, if it is stable. 

MSSM however comes with its own unpleasant baggage and must necessarily
be part of a larger more symmetric model. To get a glimpse of what this
larger model looks like, let us recall the problems that beset the MSSM.
They are the following:

\noindent (i) The MSSM symmetries allow the existence of baryon and lepton
number violating terms with arbitrary strength, a feature which not only
allows the the LSP to decay in fraction of a few years but more seriously,
it also allows the proton to decay in a fraction of a second. This is the
so called R-parity problem. For a recent review of some the consequences
of R-parity breaking, see Ref.\cite{gautam}. In view of the fact that the 
standard model led to the conservation of baryon and lepton number 
automatically (i.e. by virtue of the choice of the gauge symmetry and the
field representations), MSSM takes us a step backward.

\noindent (ii) A second problem with the MSSM lies in its predictions for 
the CP violating effects being too large. There are two
extra phases in MSSM in its most symmetric version residing in the soft 
breaking parameter $A$ and the Higgs mixing mass $\mu$. 
These phases manifest in the electric dipole moment of the neutron,
already at the one loop level leading generically to:
\begin{eqnarray}
d^e_n\simeq \frac{e}{16\pi^2}\frac{m_d}{M^4_{\tilde{q}}} 
Arg(m_{\tilde{g}} 
[A-\mu tan\beta])
\end{eqnarray}
A simple evaluation of the above down quark electric dipole moment implies
that unless either (i) the squark masses are of order 3 TeV or
(ii) $Arg({m_{\tilde{g}}A})$ and $Arg(m_{\tilde{g}}\mu)$ are less than 
$10^{-3}$
if squark masses $M_{\tilde{q}}\simeq 100 $ GeV, the edm of the neutron will
come out to be three orders of magnitude higher\cite{garisto}
than the present experimental
upper bound. In either case, we have a fine tuning problem for the
theory, the very problem supersymmetry was supposed to solve. In the first
case one has to fine tune to get the Higgs mass of order $m_W$ and in the
second case, the new phases of the model (unlike the CP phase of the standard
model) has to be tuned down by three orders of magnitude from its natural
value.

\noindent (iii) The MSSM with global R-parity conservation leads to zero mass
for the neutrinos. In view of the recent growing experimental evidences for
neutrino masses, it is more appropriate to consider extensions of the MSSM
that can lead to neutrino masses.

The simplest extension of the MSSM that solves all three of the above 
problems is the supersymmetric left-right model with the field 
content chosen to yield naturally small neutrino masses via the seesaw
mechanism. The detailed solution to the R-parity and SUSYCP problems
in the left-right symmetric models has been discussed in \cite{moh86,rasin}.
We will briefly go over these arguments in section 2 of this article, where
we also present the field content and the superpotential. 

The main focus of this article will be on the constraints on the $W_R$ scale
implied by electric charge conservation and R-parity conservation by the
ground state.

\section{The Model}

The model, which is based on the gauge group
$SU(2)_L\times SU(2)_R\times U(1)_{B-L}\times SU(3)_c$\cite{pati74}. In 
Table I, we give the particle content of the model needed to implement 
the seesaw mechanism for neutrino masses\cite{gell}. We will  suppress the 
$SU(3)_c$ indices in what follows. \\\vskip2mm

\begin{tabular}{|c|c|c|} \hline Fields  & SU$(2)_L \, \times$ 
SU$(2)_R 
\, \times$ U$(1)_{B-L}$ &  group transformation \\
                 & representation & \\ \hline Q & (2,1,$+ {1
\over 3}$) & UQ \\
$Q^c$  & (1,2,$- {1 \over 3}$) & VQ$^c$ \\ L  & (2,1,$- 1$) & UL \\
$L^c$ & (1,2,+ 1) & V$L^c$ \\
$\Phi_{1,2}$ & (2,2,0) & $U\phi V^{\dagger}$ \\
$\Delta$& (3,1,+ 2) & U$\Delta U^{\dagger}$ \\
$\bar{\Delta}$ & (3,1,$- 2$) & $U\bar{\Delta}U^{\dagger}$ \\
$\Delta^c$ & (1,3,+ 2) & $V\Delta^c V^{\dagger}$ \\
$\bar{\Delta}^c$ & (1,3,$- 2$) & $ V\bar{\Delta}^cV^{\dagger}$ \\
$S$&  (1,1,0) &  $S$ \\\hline
\end{tabular}\\ Table 1:Field content of the SUSY LR model; we 
assume that $S$
\\is odd  under parity; $U$ and $V$ denote the $SU(2)_{L,R}$ 
transformations
respectively.
\vskip 4mm

The superpotential for this theory is given by (we have suppressed 
the
generation index):         
\begin{eqnarray}  W & = & {\bf h}^{(i)}_q Q^T \tau_2 \Phi_i \tau_2 Q^c + 
{\bf
h}^{(i)}_l L^T \tau_2 \Phi_i \tau_2 L^c
\nonumber\\
  & +  & i ( {\bf f} L^T \tau_2 \Delta L + {\bf f}_c {L^c}^T \tau_2 
\Delta^c
L^c)
\nonumber\\
  & +  & M_{\Delta} [{\rm Tr} ( \Delta \bar{\Delta} ) +
 {\rm Tr} ( \Delta^c \bar{\Delta}^c )] +\lambda S(\Delta\overline{\Delta}
-\Delta^c\overline{\Delta^c}) + \mu_S S^2 \nonumber\\
 & + &
\mu_{ij} {\rm Tr} ( \tau_2 \Phi^T_i \tau_2 \Phi_j )+ W_{\it NR}
\label{eq:superpot}
\end{eqnarray}  where $W_{\it NR}$ denotes non-renormalizable terms 
arising from
higher scale physics such as grand unified theories or Planck scale 
effects.

\begin{eqnarray}
   W_{\it NR}= A [Tr(\Delta^c \overline {\Delta}^c)]^2/2  + B Tr(\Delta^c
\Delta^c) Tr (\overline{\Delta}^c \overline{\Delta}^c)/2
\end{eqnarray}  where A and B are of order 1/$M_{Planck}$ and we have 
omitted terms involving left triplet Higgs fields for the reasons stated 
below.

Our goal is to seek the ground state of this model that conserves 
electric charge and violates parity. First we note that the presence of the 
parity odd singlet
$S$ enables one to get a parity violating minimum. The effective theory 
below this scale (i.e. scale $<S>\neq 0$), can be written only in terms 
of the $\Delta^c$ and $\overline{\Delta^c}$ terms. Therefore, in what 
follows we  drop the $\Delta$ and $\overline{\Delta}$ fields.

There are now two possibilities for the vacuum state: one which 
conserves R-parity in the process of breaking the gauge symmetries. 
The Higgs vevs for this case have the following pattern:
\begin{eqnarray} <\phi>=\left(\begin{array}{cc}
\kappa & 0\\ 0 & \kappa '\end{array}\right); 
<\Delta^c>=\left(\begin{array}{cc}
0 & v_R \\ 0 & 0 \end{array}\right)
\end{eqnarray} Similar pattern for $<\overline{\Delta^c}>$ is assumed.

There is however a second possibility where in addition to the above vevs,
one could have $<\tilde{\nu^c}>\neq 0$. This ground state breaks R-parity
spontaneously. As a result, even though this does not allow the LSP to 
remain stable, baryon number remains a good symmetry and the most disastrous
limits on the R-violating couplings are avoided.

Two different bounds on the $W_R$ masses emerge for the two cases: in case
(i) where R-parity is exactly conserved, we find\cite{chacko,goran2} that
there is a lower bound on the $W_R$ mass i.e. $M_{W_R}\geq 10^{10}$ GeV;
in case (ii) on the other hand, where R-parity is spontaneously broken,
there is an upper bound on $M_{W_R}$ of less than a few TeV\cite{kuchi}.  
Below I briefly outline the main arguments leading to these bounds and
refer the reader to the original papers\cite{chacko,goran2,kuchi} for 
further details.

\section{Exact R-parity conservation and lower bound on $M_{W_R}$}

Let us first give a group theoretical argument for the existence of the
lower bound.
Using Eq. 3, we will first show that in the supersymmetric limit,
there exist two massless doubly charged superfields if we
ignore the higher dimensional terms $A$ and $B$ as well as the leptonic
couplings $f$ in the superpotential. It is easy to see that the 
superpotential in this case has a
complexified $U(3)$ symmetry (i.e. a $U(3)$ symmetry whose parameters are
taken to be complex) that operates on the $\Delta^c$ and $\bar\Delta^c$  
fields. This is due to the holomorphy of the superpotential. After one   
component of each of the above fields acquires vev as in the charge      
conserving case with $\theta=0$ (and supersymmetry
guarantees that both vev's are parallel), the resulting symmetry is the
complexified $U(2)$. This leaves 10 massless fields. Once we bring in the
D-terms and switch on the gauge fields, six of these fields become massive
as a consequence of the Higgs mechanism of supersymmetric theories.
That leaves four massless fields in the absence of higher dimensional
terms. These are the two complex
doubly charged fields. The existence of these massless particles signals
the presence of a flat direction, which has been shown to exist in this 
case\cite{kuchi1}. The existence and the  parameterization of the 
flat direction can be seen by writing down the potential in the 
supersymmetric limit:
\begin{eqnarray}
V(\Delta^c, \overline{\Delta^c})= M^2(Tr({\Delta^c}^{\dagger}\Delta^c)+
Tr({\overline{\Delta^c}}^{\dagger}\overline{\Delta^c}))\nonumber\\
+ D-terms
\end{eqnarray}
We omit the detailed form of the D-terms except to note that it is only a 
function of 
$(Tr({\Delta^c}^{\dagger}\tau_a\Delta^c)-Tr(\overline{\Delta^c}^{\dagger}
\tau_a\overline{\Delta^c}))^2$. In the limit of supersymmetry, one must have
the absolute values of the vevs of $\Delta^c$ and $\overline{\Delta^c}$ 
equal. It is then easy to see that the
flat direction can be parameterized as follows: 
\begin{eqnarray}
<\Delta^c>=v_R\left(\begin{array}{cc}
0 & sin\theta \\
cos\theta & 0 \end{array}\right)
\end{eqnarray}
Clearly $\theta=\pi/2$ corresponds to the charge conserving vacuum.

Let us now add the nonrenormalizable Planck scale induced terms to the 
superpotential. Of the two possible terms $A$ 
and $B$ given above, only the A-term has the complexified $U(3)$ symmetry. 
Hence the
supersymmetric contribution to the doubly charged particles will come only
from the B-term. It is then easy to see that if the nonrenormalizable terms
$A$ and $B$ are scaled by the Planck mass, $M_{P}$, then their 
contributions to the doubly charged fields is of order $v^2_R/M_{P}$.
Since the CERN LEP lower bound on the masses of such particles is 45 GeV,
this implies that we must have $v_R\geq 10^{10}$ GeV. 
Although the leptonic couplings do not respect 
this symmetry, they are unimportant in determinimg the vacuum structure
and hence do not effect this result.

Of course one might argue at this point that once one incorporates
supersymmetry breaking terms, the doubly charged particles might pick up
masses of order 100 GeV anyway regardless of what the value of $W_R$ is.
However, as was shown in great detail in Ref.\cite{chacko}, this does not 
happen and the bound remains as it is. Let us elaborate on this now.
The main point is that in the presence of the supersymmetry breaking terms
and in the absence of the nonrenormalizable terms the global minimum
of the potential turns out to be at $\theta=0$ as 
shown in Ref.\cite{kuchi1}, which means that electric charge is no more
respected by vacuum. This manifests itself in detailed calculation as
a negative mass-squared term for the doubly charged Higgs boson 
${\Delta^c}^{++}$. The mass-squared term is of order of the 100 GeV to a 
TeV. In 
order to have a charge conserving vacuum, we must seek a positive 
contribution of the same order to the ${\Delta^c}^{++}$ term. This is 
provided by the non-renormalizable terms which contribute an amount
$\simeq \frac{v^4_R}{M^2_{Pl}}$ to $M^2_{\Delta^c}$. It is then clear that
in order to lead to a charge conserving vacuum, we must have 
$\frac{v^2_R}{M_{Pl}}\geq 100 $ GeV leading to the lower limit on the 
right handed scale of order $10^{10}$ GeV. (We have used $M_{Pl}\simeq 
1.2\times 10^{18}$ GeV.) 

In the above discussion, we assumed that the hidden sector supersymmetry 
breaking scale is in the range of $10^{12}$ GeV or so. There are however
the scenarios for supersymmetry breaking where the hidden sector SUSY 
breaking is transmitted via the known gayuge forces\cite{rattazi}- the so 
called GMSB models where the SUSY breaking scale $\Lambda_S$ could be of 
order 100 
TeV. The $SU(2)_R$ scale in these models could therefore be higher than
$\Lambda_S$. It turns out that in these models the effective theory below 
$v_R$ contains massless doubly charged superfields. It was shown in 
\cite{chacko} that once the hidden sector SUSY breaking is turned on,
only the scalar component of the doubly charged Higgs superfield picks
up mass of order 100 GeV. Therefore one must invoke the higher 
dimensional terms $A$ and $B$ to generate enough mass for the fermionic
component. Again requiring LEP Z-decay bound of $45$ GeV for this 
particle leads to a bound of $10^{10}$ GeV for the $v_R$ scale. Thus even 
though there is no problem with charge violation by the vacuum, 
essentially the same bound on $v_R$ emerges. It is clear that the mass of 
the doubly charged superfields in both the GMSB as well as the high scale
gravity mediated models is given by $\simeq v^2_R/M_{Pl}$ and is in the
accessible range of accelerator experiments if $v_R\simeq 
10^{10}-10^{11}$ GeV. Implications for collider experiments of such a 
light doubly charged field has been extensively studied in recent 
papers\cite{dutta}.

\section{Upper limit on $M_{W_R}$ with spontaneous R-parity violation}

There is another way to lower the electric charge conserving vacuum below the
one that violates it by giving vev to the $\tilde{\nu^c}$ field as was noted
in \cite{kuchi1}. So does this mean that in the R-parity violating situation,
the $M_{W_R}$ is unrestricted ? The answer to this question is "No" since
in order to have $<\tilde{\nu^c}>\neq 0$, we must have the potential for the
field $\tilde{\nu^c}$ must have a "Mexican" hat form. Let us therefore 
look at the schematic form of the potential for the $\tilde{\nu^c}$ field.
\begin{eqnarray}
V(\tilde{\nu^c})= M^2_{\tilde{\nu^c}}\tilde{\nu^c}^{\dagger}\tilde{\nu^c}
+f~ M_{SUSY}~ v_R\tilde{\nu^c}^2 +f^2v^2_R |\tilde{\nu^c}|^2\nonumber\\
 + higher~~ powers~~ of~~ \tilde{\nu^c}+ h.c.
\end{eqnarray}
Note that if we keep the sign of the first term negative and if $fv_R\leq 
|M_{\tilde{\nu^c}}|$, then one can have $<\tilde{\nu^c}>\neq 0$. But once 
$fv_R\geq | M_{\tilde{\nu^c}}|$, the minimum of the potential under 
consideration is at $<\tilde{\nu^c}>=0$. This then means that the charge 
violating minimum becomes the lower minimum. In other words in the case
with spontaneous R-parity violation, there must be an upper limit on the
scale $v_R\leq M_{SUSY}/f$. For reasonable value of the parameters, this 
implies an upper limit on $M_{W_R}$ of at most 10 TeV's. 

\section{Testing high scale $SU(2)_R$ theories in neutrinoless double 
beta decay}

The high $SU(2)_R$ breaking scale implied by the minimal SUSYLR models 
with the seesaw mechanism with R-parity conservation decouples essentially
all new particles that are not present in the MSSM except the doubly 
charged bosons and fermions. They remain light with mass around 100 GeV
or so. Thus the low energy effective theory consists of the MSSM spectrum
with massive Majorana neutrinos and a pair of doubly charged Higgs 
superfields. If this theory is embedded into an SO(10) GUT model, then
unification of gauge coupling constants demands that the $SU(2)_R$ breaking
scale be equal to the GUT scale, $M_U$. The doubly charged fields in this 
case become superheavy and disappear from the low energy spectrum. Looking 
for the low energy 
effects of the doubly charged particles will therefore be a way to test 
between a grand unified legft-right model such as $SO(10)$ and a nonunified
SUSYLR model all the way to the Planck or string scale. 

One interesting experimental effect of the light doubly charged Higgs bosons
is in the neutrinoless double beta decay\cite{verga}. Let me explain how 
this effect 
arises and how it becomes observable despite the high $v_R$ scale. Note that
among the nonrenormalizable operators that can be added to the theory is 
the operator $\Phi\Phi\Delta^c\overline{\Delta^c}/M_{Pl}$. This leads to
a term in the potential of the form 
$\frac{M}{M_{Pl}}\phi\phi\Delta^c{\Delta^c}^{\dagger}$. Since we expect
$M\approx v_R$, the strength of this interaction is of order 
$10^{-8}$. It contributes to neutrinoless double beta decay via the
diagram in Fig.1.
\begin{figure}[htb]
\begin{center}
\epsfxsize=8.5cm
\epsfysize=8.5cm
\mbox{\hskip -1.0in}\epsfbox{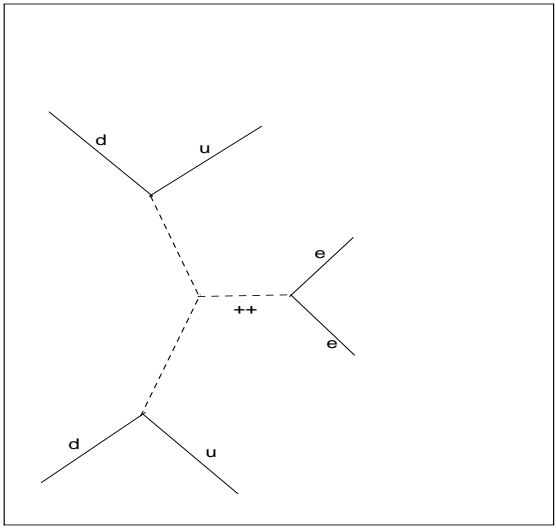}
\caption{ The Feynman diagram responsible for neutrinoless double beta decay
The top and bottom solid lines are quark lines and the middle right solid 
lines are electron lines. The dashed lines are the scalar 
bosons with appropriate quantum numbers.\label{Fig.1}} \end{center}
\end{figure}
 This leads to a double beta decay amplitude roughly of order
\begin{eqnarray}
M_{\beta\beta}\simeq \frac{g^2}{4}\left(\frac{m_d}{m_W}\right)^2 
\frac{v^2_R}{M_{Pl}}\frac{f}{M^4_{\phi}M^2_{++}}
\end{eqnarray}
The $M_{\phi}$ denotes the mass of the bidoublet Higgs field and we have
assumed that the Yukawa couplings of the bidoublet Higgs is proportional 
to the quark masses in analogy with the standard model. In principle, 
this could be bigger. Therefore our estimate is the most conservative one. 
For $v_R\simeq 10^{10}$ GeV, we find that this leads to $M_{\beta\beta}
\simeq 10^{-18}~GeV^{-5}$, which is roughly where the present 
Heidelberg-Moscow enriched Germanium limits are\cite{klap}. The amplitude
depends on the $SU(2)_R$ scale like $v^{-2}_R$ since the mass of the 
doubly charged Higgs field goes like $M_{++}\sim v^2_R$ and therefore an 
improvement in the lifetime by a factor of 100 will improve the lower limit
on $v_R$ by a factor of three.

 \section{Comments and Conclusion}

In conclusion, we have shown that in the minimal supersymmetric left-right 
model with the seesaw mechanism, the requirement that vacuum state conserve
electric charge imposes very stringent limits on the scale of the right
handed interactions. First of all there is a whole range of values for
the $SU(2)_R$ scale $v_R$ (i.e. $10^4\leq v_R/GeV \leq 10^{10}$) where
the vacuum breaks electric charge and is therefore theoretically ruled out.
If we further demand that the vacuum of the theory conserve R-parity 
automatically, then the entire range below $10^{10}$ GeV is ruled out.
On the other hand if we allow the vacuum to break R-parity, then the range
above 10 TeV is ruled out. It is interesting to note that the higher mass 
range seems to be preferred by the conventional neutrino mass schemes
being discussed in the literature. It is also important to point out that
the lower mass range can be substantially covered by the proposed GENIUS 
double beta decay experiment\cite{klap} as well as the ATLAS detector in 
the LHC experiment\cite{coolot}

I wish to thank Z. Chacko, B. Dutta and G. Senjanovi\'c  for many 
discussions on the topics covered in this talk and Pran Nath for hospitality 
during the PASCOS98 workshop. This work is supported by the National 
Science foundation grant no. PHY- 9802551.

\end{document}